\documentclass{pasj01}
\draft 
\Received{2018 May 9}
\Accepted{2018 November 23}
\Published{$\langle$publication date$\rangle$}

\begin{document}

\title{Suzaku detection of enigmatic geocoronal solar wind 
charge exchange event associated with coronal mass ejection}
\author{Daiki \textsc{Ishi},\altaffilmark{1,}$^{*}$
Kumi \textsc{Ishikawa},\altaffilmark{2}
Masaki \textsc{Numazawa},\altaffilmark{1}
Yoshizumi \textsc{Miyoshi},\altaffilmark{3} 
Naoki \textsc{Terada},\altaffilmark{4}
Kazuhisa \textsc{Mitsuda},\altaffilmark{2}
Takaya \textsc{Ohashi},\altaffilmark{1} and
Yuichiro \textsc{Ezoe}\altaffilmark{1}}
\altaffiltext{1}{Tokyo Metropolitan University, 1-1 Minami-Osawa, Hachioji, Tokyo 
192-0397, Japan}
\altaffiltext{2}{Institute of Space and Astronautical Science, Japan Aerospace Exploration Agency, 
3-1-1 Yoshinodai, Chuo-ku, Sagamihara, Kanagawa 252-5210, Japan}
\altaffiltext{3}{Nagoya University, Furo-cho, Chikusa-ku, Nagoya, Aichi 464-8601, Japan}
\altaffiltext{4}{Tohoku University, 6-3 Aramaki-Aza-Aoba, Aoba-ku, Sendai, Miyagi 980-8578, Japan}
\email{ishi-daiki@ed.tmu.ac.jp}


\maketitle

\begin{abstract}
Suzaku detected an enhancement of soft X-ray background associated with solar 
eruptions on 2013 April 14--15.
The solar eruptions were accompanied by an M6.5 solar flare and a coronal mass 
ejection with magnetic flux ropes. 
The enhanced soft X-ray background showed a slight variation in half a 
day and then a clear one in a few hours. 
The former spectrum was composed of oxygen emission lines, while the 
later one was characterized by a series of emission lines from highly ionized carbon to 
silicon.
The soft X-ray enhancement originated from geocoronal solar wind 
charge exchange.
However, there appeared to be no significant time correlation with the 
solar wind proton flux measured by the ACE and WIND satellites.
From other solar wind signatures, we considered that an interplanetary 
shock associated with the coronal mass ejection and a turbulent sheath immediately 
behind the shock compressed the ambient solar wind ions and then resulted in the soft 
X-ray enhancement.
Furthermore, the enriched emission lines were presumed to be due to an 
unusual set of ion abundances and ionization states within the coronal mass ejection.
We found a better time correlation with the solar wind alpha flux rather than the solar 
wind proton flux. 
Our results suggest that the solar wind proton flux is not always a good indicator of 
geocoronal solar wind charge exchange, especially associated with 
coronal mass ejections. 
Instead, the solar wind alpha flux should be investigated when such a 
soft X-ray enhancement is detected in astronomical observations. 
\end{abstract}

\section{Introduction}

Solar eruptions such as solar flares and coronal mass ejections (CMEs) are powerful 
explosive phenomena in our solar system. 
Solar flares, intense bursts of radiation across the entire electromagnetic spectrum 
ranging from radio waves to $\gamma$-rays, release more than 10$^{32}$ erg of 
magnetic energy in tens of minutes (e.g., \cite{for02}).
The CMEs are huge clouds of coronal magnetized plasma with a typical mass of 
10$^{15-16}$ g and speeds of 250--1000 km s$^{-1}$ into interplanetary space 
\citep{web12}.
These eruptions are more frequent during the active phase of the solar cycle and have 
a significant influence on planetary atmospheres and their surrounding environments. 

Solar activity influences X-ray astronomical observations by spacecrafts. 
The spacecrafts often detect increased background X-rays when their lines of sight 
directions face on the sunlit side of the Earth's atmosphere (e.g., \cite{mck82}). 
X-rays from the sunlit atmosphere are due to Thomson scattering of solar X-rays by 
electrons in atmospheric atoms and molecules, as well as absorption of incident 
solar X-rays followed by the emission of characteristic K lines of atmospheric 
neutrals.
These background X-rays are usually removed 
by discarding time intervals when the line of sight direction faces on 
the sunlit atmosphere. 

The other important background source is solar wind charge exchange (SWCX) in 
the Earth's exosphere or geocoronal SWCX. 
A highly charged ion in the solar wind strips an electron from an 
exospheric neutral atom or molecule, and then releases X-ray 
or ultraviolet photons when the electron cascades into the ground state. 
The most explicit signs of the geocoronal SWCX were discovered 
during the ROSAT All Sky Survey (e.g., \cite{sno94, cra01}).
Thanks to recent X-ray astronomical observations with X-ray CCDs onboard Chandra, 
XMM-Newton, and Suzaku (e.g., \cite{sno04,warg04,fuj07,car08,ezo10,ezo11,ish13}), 
the geocoronal SWCX are now established as time variable diffuse background.
Careful checks of background X-rays combined with simultaneously observed solar 
wind data are essential for removing the geocoronal SWCX.

Geocoronal SWCX provides not only background signals but also important 
information such as the exospheric density and the minor constituents of the 
solar wind. 
Models to simulate the spatial distribution of the geocoronal SWCX (e.g., \cite{rob06}) 
indicated that the distribution is non-uniform and higher at the dayside of the Earth's 
magnetosphere, especially the magnetosheath and the magnetosheric cusps. 

\citet{car08} and \citet{car11} systematically searched for geocoronal SWCX 
emission from XMM-Newton archival data. 
Most of the observations affected by the geocoronal SWCX were found through the 
sub-solar side of the magnetosheath. 
The strong SWCX emission was sometimes observed even when the line of sight 
direction did not intersect the magnetosheath. 
They mentioned that the latter cases probably originated from 
non-geocoronal SWCX in the heliosphere (e.g., \cite{cra00,kou07}).
\citet{car10} argued that the most spectrally rich case was attributed 
to a CME passing through the Earth on 2001 October 21.

The X-ray Imaging Spectrometer (XIS) onboard Suzaku consists of 
three front-illuminated (FI) CCDs and one back-illuminated (BI) CCD and has a 
good energy response and a low background rate \citep{mit07,koy07}.
\citet{fuj07} and \citet{ezo11} discovered geocoronal SWCX events in the directions 
of the magnetospheric cusps, while \citet{ezo10} found an event toward the sub-solar 
side of the magnetosheath. 
\citet{ish13} detected one of the strongest geocoronal SWCX emission when the line 
of sight direction faced on the anti-sunward side of the magnetosheath. 
These events showed a significant correlation of the X-ray flux with the simultaneously 
observed solar wind proton and ion fluxes.

Motivated by these Suzaku detections and their correlations with the 
solar wind data, we are proceeding a systematic search for the geocoronal SWCX 
events from all the Suzaku archival data.
During this systematic search, we found a distinctive geocoronal SWCX event most 
probably associated with an M6.5 solar flare and a CME recorded on 2013 April 11. 
This event could be missed in the standard search for the geocoronal SWCX events 
using a time correlation with the solar wind proton flux. 
In this paper, we describe characteristics of this geocoronal SWCX event.

\section{Observation}

Suzaku observed a supernova remnant (SNR) 0509$-$67.5 in the Large Magellanic 
Cloud (LMC) on 2013 April 11--15 (ObsID: 508072010). 
The pointing center was (RA, Dec)$_{\rm J2000.0}$ $=$ (\timeform{77D.393}, 
\timeform{-67D.525}), which corresponds to ($l$, $b$) $=$ (\timeform{278D.147}, 
\timeform{-34D.586}) in Galactic coordinates. 
This SNR has a shell-like structure extending to the angular radius of \timeform{14''.8} 
in the X-ray band \citep{warr04}. 
Its X-ray spectrum consists of a non-thermal continuum component and many bright 
emission lines originating from thermal plasma such as silicon, sulfur, and iron. 
\citet{yam14} analyzed all extant Suzaku data including this observation and constrained 
their progenitor types. 

\begin{figure}[t]
\begin{center}
\includegraphics[width=1.0\textwidth]{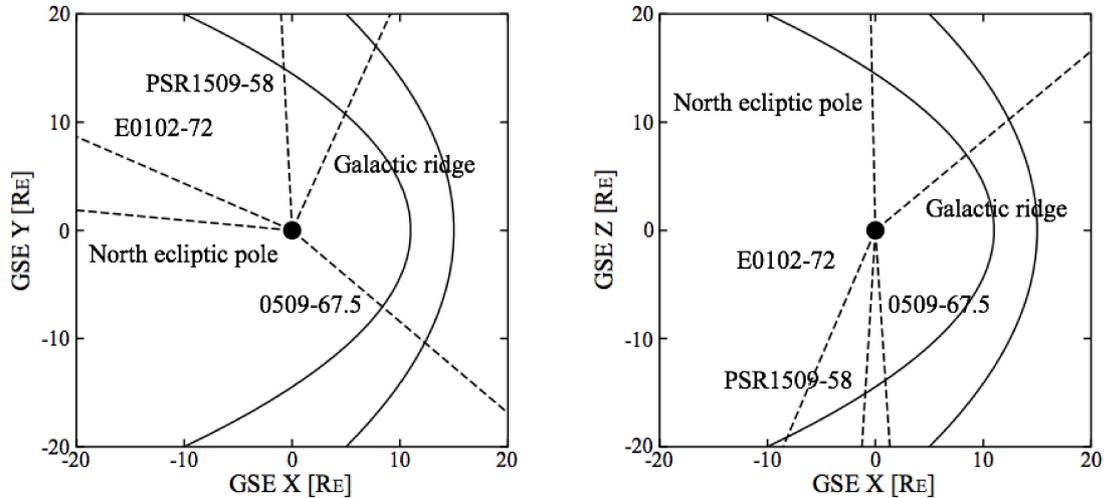} 
\end{center}
\caption{
Average lines of sight directions in the GSE XY and XZ planes during this observation 
(0509$-$67.5), the north ecliptic pole \citep{fuj07}, the galactic ridge \citep{ezo10}, 
PSR B1509$-$58 \citep{ezo11}, and 1E 0102.2$-$7219 \citep{ish13}. 
Filled black circles in each panels represent Earth. 
Two solid curves indicate approximate positions of the Earth's magnetopause and 
bow shock.
}
\label{fig1} 
\end{figure}

We utilized the same data for an investigation of the geocoronal SWCX emission. 
Figure \ref{fig1} shows the average line of sight direction in Geocentric Solar Ecliptic 
(GSE) coordinates during this observation. 
A normalized pointing vector in GSE coordinates was (0.0668, $-$0.0564, $-$0.9958).
It passed through the southern magnetospheric cusp where stronger SWCX flux would 
be expected from the past Suzaku observation of the geocoronal SWCX event on 2005 
August 23--24 \citep{ezo11}.
This observation covered from 2013 April 11, 01:26 to April 15, 03:01 UT (Day of Year, 
DOY, 101.06--105.13 in 2013). 

\begin{figure}[p] 
\begin{center}
\includegraphics[width=0.6\textwidth]{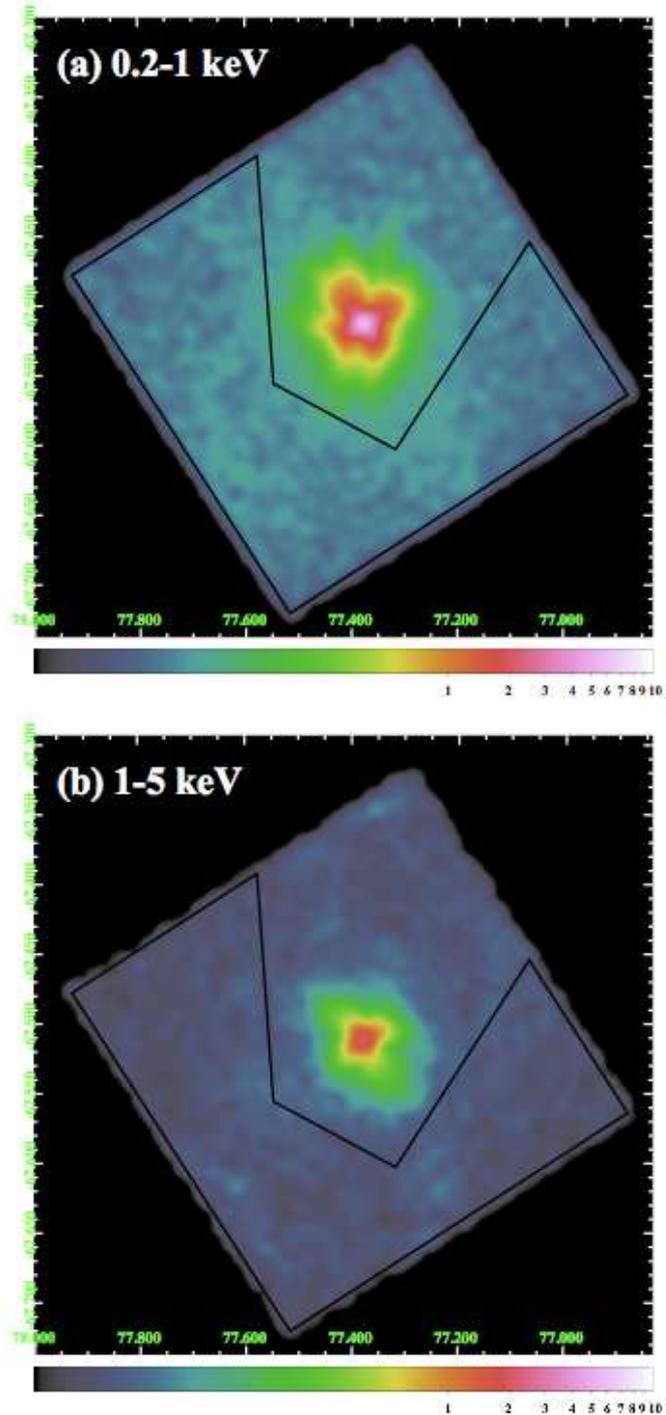} 
\end{center}
\caption{
XIS images in (a) 0.2--1 keV (XIS1) and (b) 1--5 keV (XIS3) displayed on J2000.0 
coordinates. 
For clarity, the images were smoothed by a Gaussian profile of $\sigma$ = 15 pixels 
corresponding to $\sim$\timeform{15''}. 
A color scale refers to counts per pixel. 
The polygon region is used for light curve and spectral analyses.
}
\label{fig2} 
\end{figure}

We started data analyses from cleaned event data that had already been 
pre-processed with the standard screening criteria.\footnote{$\langle$https://heasarc.gsfc.nasa.gov/docs/suzaku/analysis/abc/abc.html$\rangle$.}
Its processing version was 3.0.22.44. 
The standard screening removes high-background events mainly during passages 
through the South Atlantic Anomaly, and selects good time intervals when the satellite 
can stably point at the astronomical source without the Earth occultation. 
We utilized the HEAsoft version 6.21 package to handle scientific products and 
instrumental response files. 
The XIS2 turned off due to damages by a micro-meteorite on 2006 November 
9.\footnote{$\langle$http://www.astro.isas.jaxa.jp/suzaku/news/2006/1123/$\rangle$.}
The XIS0 was also damaged by a similar accident on 2009 June 23.\footnote{$\langle$http://www.astro.isas.jaxa.jp/suzaku/news/2009/0702/$\rangle$.}
Hence, we used only XIS1 and 3 data in this paper.
The XIS1 has higher sensitivity to soft X-rays because it is a BI-type CCD and
the XIS3 is a FI-type one. 
Figure \ref{fig2} shows XIS images in two representative energy bands. 
To minimize contamination from the X-ray emission of the main target located at the 
center of the field of view, we defined a polygon region and called it a terrestrial 
diffuse X-ray (TDX) region whose total area is 175.1 arcmin$^2$. 

An additional screening was necessary for the cleaned data so as to avoid 
contamination by scattering of solar X-rays from the Earth's atmosphere 
(e.g., \cite{ezo11,sek14}). 
We found neutral nitrogen and oxygen emission lines in 
the TDX spectrum when the elevation angle from the Earth rim (ELV) was 
\timeform{5D}--\timeform{10D}, and then screened the data with the 
ELV $>$\timeform{10D} criteria.
After this screening, these emission lines were negligible and an effective 
exposure time was 157.4 ks. 

\section{Light Curve}

\begin{figure}[t] 
\begin{center}
\includegraphics[width=1.0\textwidth]{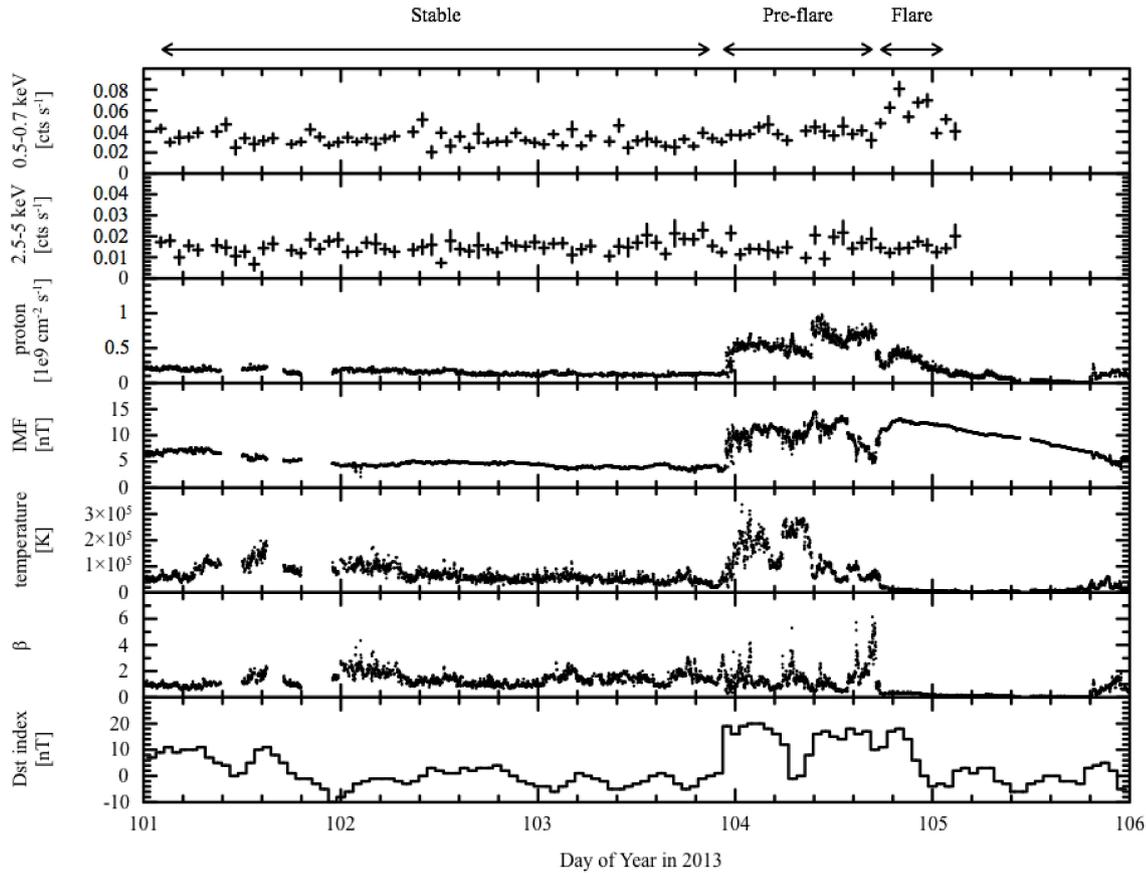} 
\end{center}
\caption{
XIS light curves in 0.5--0.7 keV (XIS1) and 2.5--5 keV (XIS3), solar wind proton flux, 
interplanetary magnetic fields frozen-in solar wind plasma, solar wind proton temperatures, 
solar wind plasma beta, and Dst index as function of DOY in 2013. 
The vertical errors are 1$\sigma$ significance. 
The sole wind parameters were taken from the high-resolution OMNI data time-shifted 
to the Earth's bow shock nose. 
The Dst index was taken from the World Data Center for Geomagnetism, Kyoto. 
}
\label{fig3} 
\end{figure}

We first extracted X-ray light curves from the TDX region in the 0.5--0.7 and 
2.5--5 keV bands as shown in figure \ref{fig3}. 
These light curves are binned with a time bin of 4096 s. 
The soft X-ray band could be affected by oxygen emission lines often seen in the 
geocoronal SWCX events, while the harder one is almost composed of a 
non-SWCX continuum. 
The 0.5--0.7 keV count rate shows a sudden enhancement just before the end of this 
observation. 
Furthermore, there slightly appears to be an increase before the sudden 
enhancement.
Hereafter, we define the ``stable'' period as DOY 101.06--103.89, the ``pre-flare'' period 
as DOY 103.98--104.71, and the ``flare'' period as DOY 104.78--105.07.
The average rate in the ``stable'' period is 0.033 $\pm$ 0.006 cts s$^{-1}$, while those 
in the ``pre-flare'' and ``flare'' periods are 0.039 $\pm$ 0.005 and 0.061 $\pm$ 0.013 
cts s$^{-1}$, respectively.
Errors are 1$\sigma$ significance. 
The count rate increased slightly during the ``pre-flare'' period and by a factor of 
$\sim$2 from the ``stable'' period to the ``flare'' period.
In contrast, the 2.5--5 keV count rate shows less variability with the average rates of 
0.015 $\pm$ 0.003 cts s$^{-1}$ in the ``stable'' period, 0.015 $\pm$ 0.004 cts s$^{-1}$ 
in the ``pre-flare'' period, and 0.014 $\pm$ 0.002 cts s$^{-1}$ in the ``flare'' period.

We then plotted four solar wind parameters: the solar wind proton flux, 
the interplanetary magnetic field (IMF) frozen-in the solar wind plasma, the solar wind 
proton temperature, and the solar wind plasma beta (the ratio of plasma to magnetic 
pressure).
These parameters were taken from high-resolution OMNI data time-shifted to the Earth's 
bow shock nose.\footnote{$\langle$https://omniweb.gsfc.nasa.gov/ow\_min.html$\rangle$.}
The solar wind proton flux suddenly increases just before DOY 104, and gradually 
decreases around DOY 104.8.
The IMF exhibits two features in response to the sudden increase and the gradual 
decrease.
One is the intense fluctuation of enhanced magnetic fields between DOY 104 and 
104.7.
The other is the strong and smooth magnetic field after DOY 104.8.
The solar wind proton temperature also rises up with the discontinuous change in the 
solar wind proton flux, and then becomes extremely low at the same time as the solar 
wind plasma beta is largely depressed.
These solar wind signatures are suggestive of a CME-induced interplanetary shock and 
the passage of a magnetic cloud that is a subset of CMEs.

Magnetic clouds are well defined observationally as an interplanetary 
structure possessing the following characteristics: enhanced magnetic fields that 
smoothly rotate, low proton temperatures, and low plasma beta (e.g., \cite{bur81, zur06}).
An M6.5 solar flare was recorded on April 11, 06:55 UT (DOY 101.29).
The CME was almost simultaneously erupted with a fast speed of $\sim$800 km s$^{-1}$.
The ejected CME drove an interplanetary shock and formed a turbulent sheath 
immediately behind the shock.
Highly compressed solar wind in the turbulent sheath results in the discontinuous change 
and the subsequent fluctuation just before DOY 104.
More subsequently, the CME or the magnetic cloud arrived at Earth around DOY 104.8.

We also plotted hourly averages of the Dst index provided by the World Data Center 
for Geomagnetism, Kyoto, Japan.\footnote{$\langle$http://wdc.kugi.kyoto-u.ac.jp/dstae/index.html$\rangle$.}
The Dst index is a measure of geomagnetic disturbances based on averaging the 
horizontal geomagnetic component from mid-latitude and equatorial magnetometer 
data.
Negative values of the Dst index indicate that a geomagnetic storm is in 
progress with the growth of magnetospheric ring current (e.g., \cite{gon94, miy05}).
Such negative values are seeable when the IMF remains southward directions for a 
prolonged time.
However, the IMF was northward within the magnetic cloud and there were no dramatic 
decreases of the Dst index during this observation.
Instead, the Dst index shows positive values after the arrival of the CME-induced 
interplanetary shock and the passage of the turbulent sheath.
These positive values result from compression of the magnetosphere and the consequent 
approach of magnetopause current.

From comparison between the X-ray light curve and the solar wind 
parameters, the X-ray enhancement in the ``pre-flare'' and ``flare'' periods should be 
closely related to the passage of the turbulent sheath associated with the CME-induced 
interplanetary shock and the arrival of the CME.
The solar wind proton flux in the ``pre-flare'' period is higher than that in the ``flare'' period.
However, the X-ray enhancement in the ``pre-flare'' period is less remarkable than 
that in the ``flare'' period.
Therefore, a correlation analysis between the X-ray light curve and the solar wind proton 
flux would miss the geocoronal SWCX events such as this one.

\begin{figure}[p] 
\begin{center}
\includegraphics[width=1.0\textwidth]{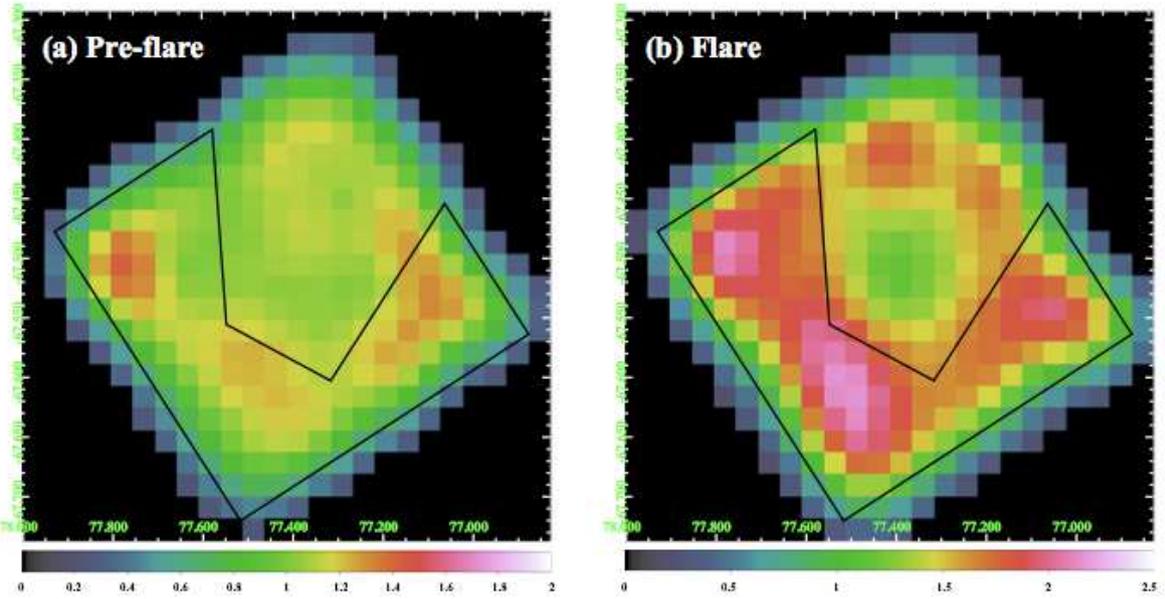} 
\end{center}
\caption{
XIS1 0.5--0.7 keV ratio maps in the (a) ``pre-flare'' and (b) ``flare'' periods. 
Each images are divided by that in the ``stable'' period. 
The difference of exposure times is corrected. 
For clarity, the images are binned by 64 pixels. 
The color scale refers to the ratio. 
The polygon region marks the TDX region.
}
\label{fig4} 
\end{figure}

\begin{figure}[p] 
\begin{center}
\includegraphics[width=1.0\textwidth]{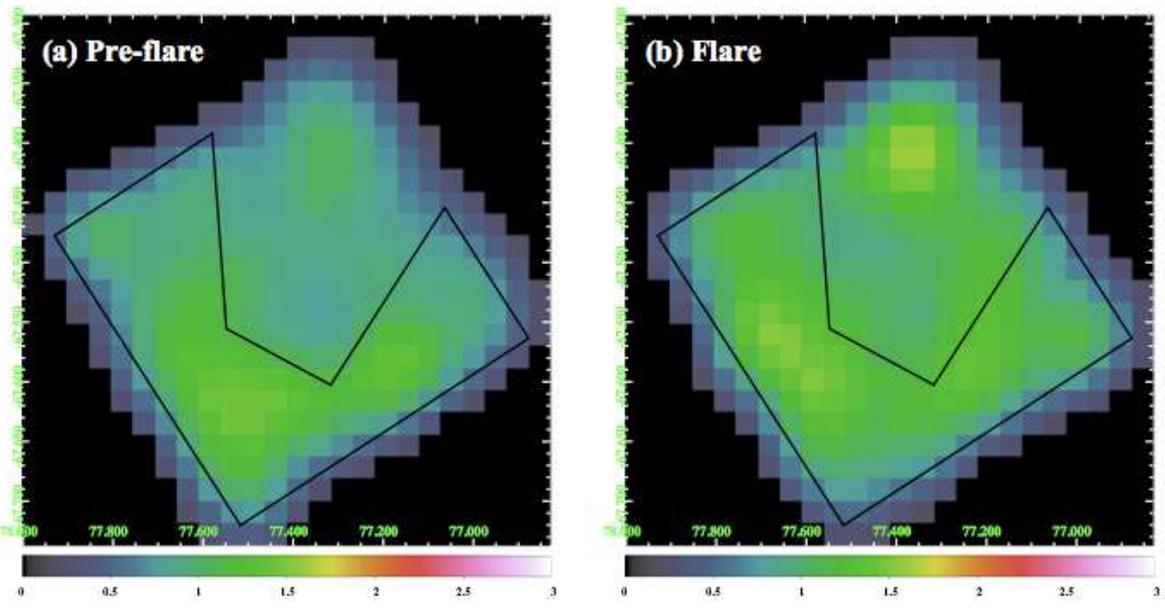} 
\end{center}
\caption{
XIS3 2.5--5 keV ratio maps in the (a) ``pre-flare'' and (b) ``flare'' periods. 
Manners are the same as in figure \ref{fig4}.
}
\label{fig5} 
\end{figure}

To check another possibility, i.e., whether the X-ray enhancement arises from leaked 
photons of the bright X-ray source and/or time variability of any other point sources, 
we compared images of  the ``pre-flare'' and ``flare'' periods with that of 
the ``stable'' period. 
Figure \ref{fig4} shows ratio maps produced by dividing the ``pre-flare'' and ``flare'' 
periods by the ``stable'' period in the 0.5--0.7 keV band.
The ratio of the TDX region increases slightly in the ``pre-flare'' period and by a factor 
of $\sim$2 in the ``flare'' period.
In contrast, the central bright X-ray source is rather steady.
Hence, the leaked photons are negligible in the TDX region.
The time variability of any other point sources also contributes little to the increased ratio 
in the entire of the TDX region.
Figure \ref{fig5} shows ratio maps in the 2.5--5 keV band. 
The entire of the TDX region is almost steady during the ``pre-flare'' and ``flare'' periods.

\section{Spectrum}

The majority of SWCX emission lines have been found in the soft X-ray band below 2 keV, 
where the BI-type CCD has higher sensitivity than the FI-type one. 
We thus focused on the XIS1 spectra extracted from the TDX region 
during the ``stable'', ``pre-flare'', and ``flare'' periods. 
These spectra included the instrument and sky backgrounds. 
The instrumental background rate was almost constant due to the 
low-Earth orbit.
The sky background originates from diffuse galactic and 
extragalactic emission and its spectral feature does not vary temporally. 
Therefore, we assumed the background spectra as constant components during this 
observation in the same way as \citet{ezo11}. 

\begin{figure}[t] 
\begin{center}
\includegraphics[width=1.0\textwidth]{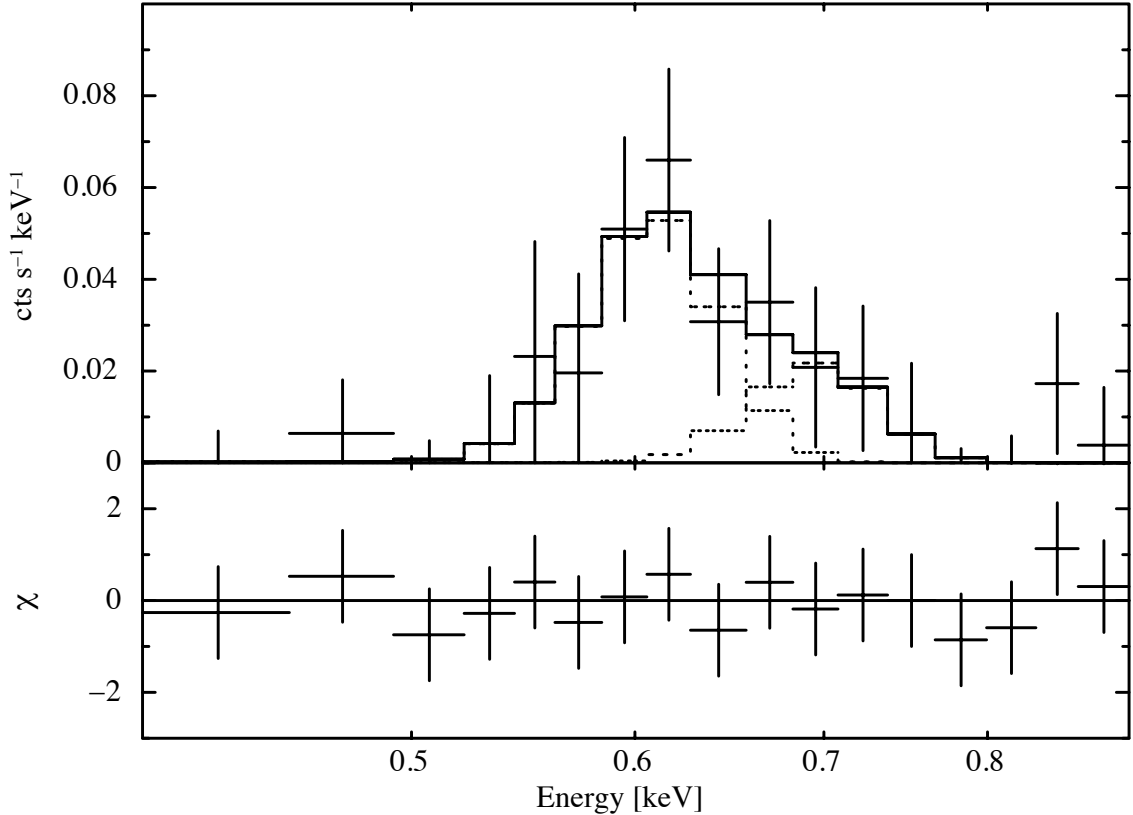} 
\end{center}
\caption{
XIS1 spectrum in the ``pre-flare'' period. 
The ``stable'' spectrum is subtracted as a background. 
The two Gaussian model is used and their parameters are listed in table \ref{tab1}. 
}
\label{fig6} 
\end{figure}

\begin{table}[t]
\tbl{Result of two Gaussian fits to the spectrum shown in figure \ref{fig6}.\footnotemark[$*$]}{%
\begin{tabular}{ccccl} \hline
Model & $E_c$\footnotemark[$\dagger$] & Normalization\footnotemark[$\ddagger$] & $f_X$\footnotemark[$\S$] & Line identification \\ \hline
1 & 608 $\pm$ 18 eV & 6.6 $^{+3.0}_{-2.5}$ & 9.6 $\times$ 10$^{-14}$ & O\,\emissiontype{VII} (f 561 eV, i 569 eV, r 574 eV)\footnotemark[$\|$] \\
2 & 694 $^{+45}_{-52}$ eV & 1.8 $\pm$ 1.5 & 2.9 $\times$ 10$^{-14}$ & O\,\emissiontype{VIII} (653 eV) \\ \hline
$\chi^2$/d.o.f & 4.78/13 & & & \\ \hline
\end{tabular}}
\label{tab1}
\begin{tabnote}
\footnotemark[$*$] All the line widths are fixed at 0 eV. \\
\footnotemark[$\dagger$] $E_c$ is the line center energy. \\
\footnotemark[$\ddagger$] Normalization is in units of 
photons s$^{-1}$ cm$^{-2}$ str$^{-1}$. \\
\footnotemark[$\S$] $f_X$ is the energy flux in erg s$^{-1}$ cm$^{-2}$. \\
\footnotemark[$\|$] f, i, and r denote forbidden, intercombination, and resonance lines. 
\end{tabnote}
\end{table}

Figure \ref{fig5} shows the spectrum produced by subtracting the ``stable'' 
period from the ``pre-flare'' period. 
The spectrum represents a spectral change during the ``pre-flare'' period. 
The data was modeled with two narrow Gaussians. 
These parameters are summarized in table \ref{tab1}. 
The line center energies were consistent with oxygen emission lines (0.5--0.7 keV) often 
seen in the typical geocoronal SWCX spectra. 
This indicates that highly ionized oxygen within the ambient solar wind increases in the 
turbulent sheath and then produces the geocoronal SWCX emission.

\begin{figure}[t] 
\begin{center}
\includegraphics[width=1.0\textwidth]{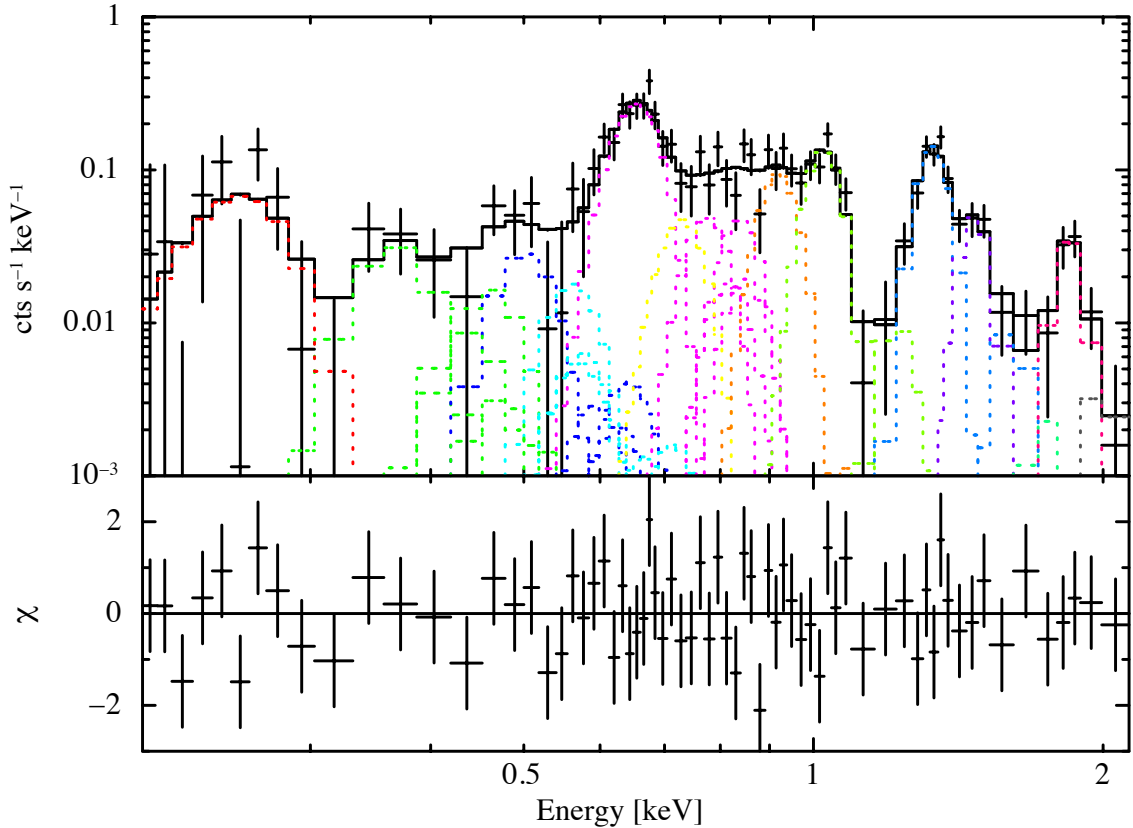} 
\end{center}
\caption{
XIS1 spectrum in the ``flare'' period. 
The ``stable'' spectrum is subtracted as a background. 
The SWCX model is used and their parameters are listed in table \ref{tab2}. 
Emission lines are color coded: C\,\emissiontype{VI} (green), 
N\,\emissiontype{VII} (blue), O\,\emissiontype{VII} (light-blue), 
O\,\emissiontype{VIII} (magenta), Fe\,\emissiontype{XVII} (yellow), 
Ne\,\emissiontype{IX} (orange), Ne\,\emissiontype{X} (yellow-green), 
Mg\,\emissiontype{XI} (blue-cyan), Mg\,\emissiontype{XII} (blue-magenta), 
Al\,\emissiontype{XIII} (green-cyan), Si\,\emissiontype{XIII} (red-magenta), 
and Si\,\emissiontype{XIV} (dark-gray). 
}
\label{fig7} 
\end{figure}

\begin{table}[t]
\tbl{Best-fit parameters of the SWCX model shown in figure \ref{fig7}.\footnotemark[$*$]}{%
\begin{tabular}{cccc} \hline
Ion & Principal energy [eV] & Normalization & $f_X$ \\ \hline
C band lines & 253 $^{+17}_{-19}$  & 24 $^{+23}_{-15}$ & 1.4 $\times$ 10$^{-13}$ \\ 
C\,\emissiontype{V} & 299 & -- & -- \\ 
C\,\emissiontype{VI} & 367 & 19 $\pm$ 9 & 1.7 $\times$ 10$^{-13}$ \\ 
N\,\emissiontype{VI} & 420 & -- & -- \\ 
N\,\emissiontype{VII} & 500 & 4.6 $\pm$ 3.8 & 5.8 $\times$ 10$^{-14}$ \\
O\,\emissiontype{VII} & 561 & 5.4 $\pm$ 5.0 & 7.3 $\times$ 10$^{-14}$ \\
O\,\emissiontype{VIII} & 653 & 33 $\pm$ 3 & 5.4 $\times$ 10$^{-13}$ \\
Fe\,\emissiontype{XVII} & 730 & 3.4 $\pm$ 1.9 & 5.8 $\times$ 10$^{-14}$ \\
Fe\,\emissiontype{XVII} & 820 & -- & -- \\
Fe\,\emissiontype{XVIII} & 870 & -- & -- \\
Ne\,\emissiontype{IX} & 920 & 4.5 $\pm$ 1.1 & 9.8 $\times$ 10$^{-14}$ \\
Fe\,\emissiontype{XX} & 960 & -- & -- \\
Ne\,\emissiontype{X} & 1022 & 5.6 $\pm$ 0.9 & 1.4 $\times$ 10$^{-13}$ \\
Ne\,\emissiontype{IX} & 1100 & 0.05 ($<0.62$) & 1.3 $\times$ 10$^{-15}$ \\
Ne\,\emissiontype{X} & 1220 & 0.39 ($<0.89$) & 1.1 $\times$ 10$^{-14}$ \\
Mg\,\emissiontype{XI} & 1330 & 5.3 $\pm$ 0.7 & 1.7 $\times$ 10$^{-13}$ \\
Mg\,\emissiontype{XII} & 1470 & 1.8 $\pm$ 0.5 & 6.4 $\times$ 10$^{-14}$ \\
Mg\,\emissiontype{XI} & 1600 & 0.50 $\pm$ 0.45 & 1.9 $\times$ 10$^{-14}$ \\
Al\,\emissiontype{XIII} & 1730 & 0.12 ($<0.58$) & 5.0 $\times$ 10$^{-15}$ \\
Si\,\emissiontype{XIII} & 1850 & 1.7 $\pm$ 0.5 & 7.5 $\times$ 10$^{-14}$ \\ 
Si\,\emissiontype{XIV} & 2000 & 0.22 ($<0.62$) & 1.0 $\times$ 10$^{-14}$ \\ \hline
$\chi^2$/d.o.f & 52.64/53 & & \\ \hline
\end{tabular}}
\label{tab2}
\begin{tabnote}
\footnotemark[$*$] Definitions of parameters are the same as in table \ref{tab1}. \\
\end{tabnote}
\end{table}

We then subtracted the spectrum of the ``stable'' period from that of the 
``flare'' period.
Figure \ref{fig7} shows the spectrum after this subtraction. 
The spectrum clearly contained prominent oxygen emission lines and 
some excess emission lines above 0.7 keV. 
A series of the excess emission lines are constituted by enriched 
solar wind minor ions such as neon, magnesium, and silicon. 

For the emission lines from highly ionized carbon, nitrogen, and oxygen, we 
used a theoretical SWCX emission-line model constructed by \citet{bod07}.
This model shows the relative emission cross-sections of highly charged ions 
(C\,\emissiontype{V}, C\,\emissiontype{VI}, N\,\emissiontype{VI}, N\,\emissiontype{VII}, 
O\,\emissiontype{VII}, and O\,\emissiontype{VIII}) in collision with atomic hydrogen for 
several solar wind velocities. 
We utilized the values for the velocity of 400 km s$^{-1}$, which was close to the average 
velocity of $\sim$450 km s$^{-1}$ during the ``flare'' period. 
The normalization of the principal transitions from each ion was fitted as a free parameter, 
whereas those of the minor transitions were fixed with respect to each principal transition. 

The best-fit parameters are summarized in table \ref{tab2}. 
Similar to \citet{ezo11}, we added an extra Gaussian to reproduce the 
lowest energy emission line around 0.25 keV.
At higher energies than the oxygen emission lines, we also added fourteen narrow 
Gaussians detected by \citet{car10}. 
The normalizations of C\,\emissiontype{V}, N\,\emissiontype{VI}, 
Fe\,\emissiontype{XVII}, Fe\,\emissiontype{XVIII}, and Fe\,\emissiontype{XX}
were zero and negligible.
We thus excluded these transitions form the best-fit parameters.

\begin{figure}[t] 
\begin{center}
\includegraphics[width=1.0\textwidth]{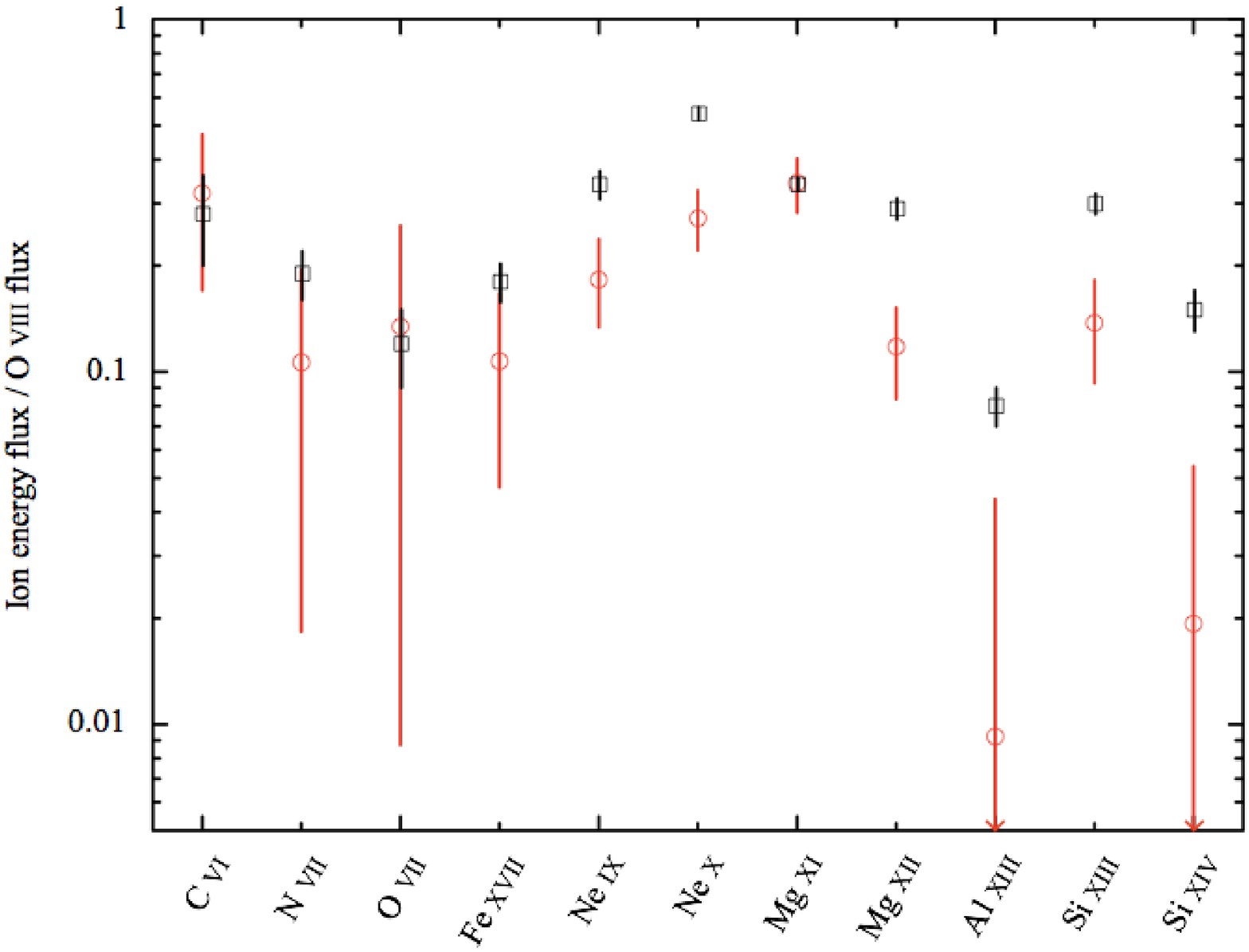} 
\end{center}
\caption{
Line energy flux ratio to the O\,\emissiontype{VIII}.
Red and black data points indicate the best-fit parameters of this observation and 
\citet{car10}, respectively.
}
\label{fig8} 
\end{figure}

The number of the detected emission lines was one of the richest case in the past 
geocoronal SWCX events. 
We succeeded in detecting the various emission lines from highly ionized 
carbon to silicon. 
Figure \ref{fig8} shows the energy flux ratios of the SWCX ion lines to the 
O\,\emissiontype{VIII} line in comparison with the XMM-Newton 
observation of the CME-induced geocoronal SWCX event reported by \citet{car10}, 
which was another case consisting rich emission lines. 
The ion composition during this observation showed the same tendency of the 
XMM-Newton observation. 
This supports that the spectral change during the ``flare'' period was 
related to the arrival of the CME.

The spectra during the ``pre-flare'' and ``flare'' periods are potentially 
influenced by the hard continuum originating from particle background (e.g., soft 
protons as shown in \cite{car10}). 
We checked these spectra after the subtraction of the ``stable'' period 
and found no significant excess component in the 2.5--5 keV band.
Therefore, the particle continuum was negligible below 2 keV.

\section{Time Correlation}

The geocoronal SWCX emission is expected to be proportional to the 
flux of highly charged ions in the solar wind and the column density of neutral atoms 
in the Earth's exosphere. 
The solar wind ion flux is usually replaced with the solar wind proton flux when the 
solar wind ion to proton abundance ratio remains constant. 
This assumption is acceptable for the ambient solar wind, but not for the CMEs. 
We thus need to refer other information about the solar wind during this observation.

\begin{figure}[t] 
\begin{center}
\includegraphics[width=1.0\textwidth]{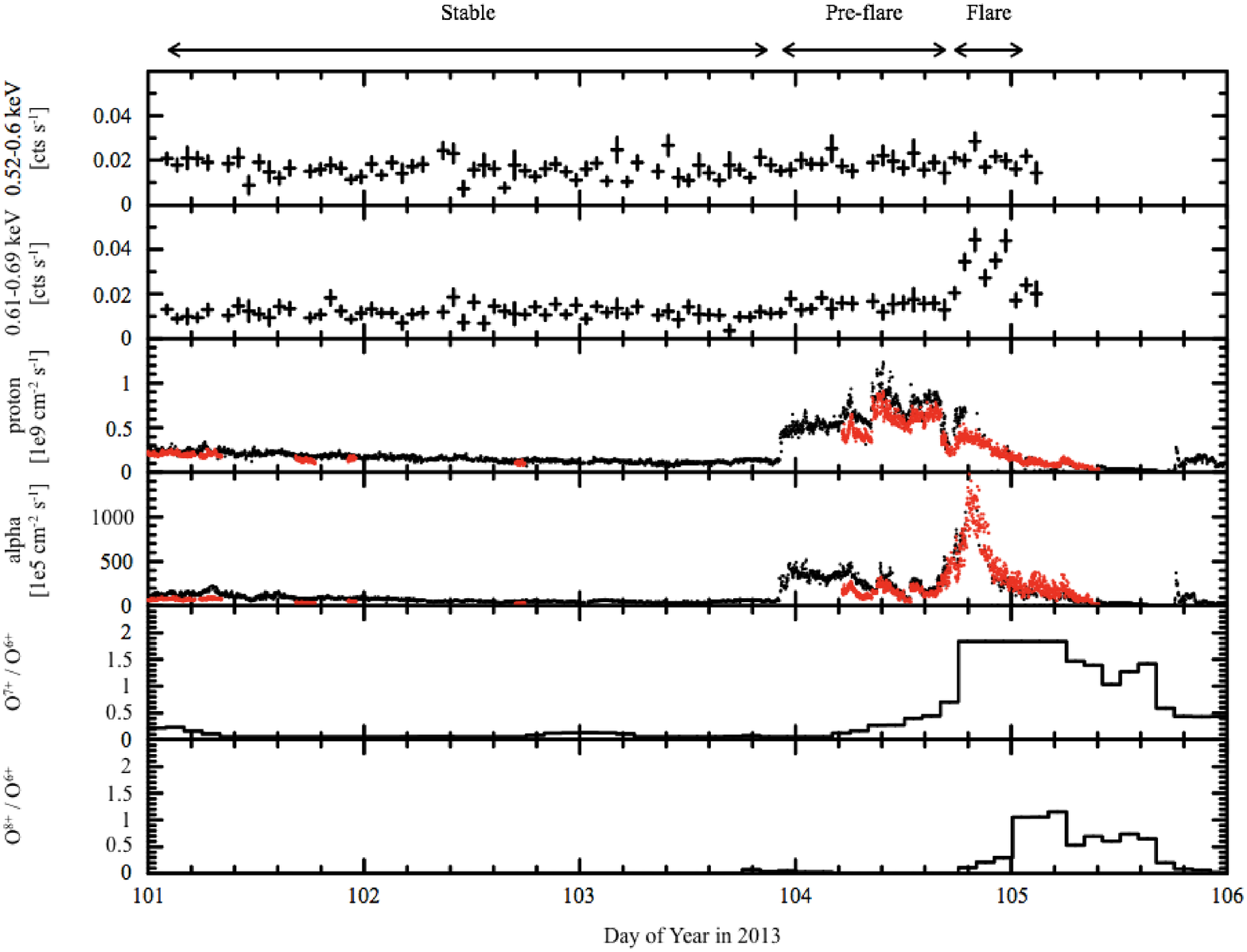} 
\end{center}
\caption{
XIS1 light curves in 0.52--0.6 keV (O\,\emissiontype{VII}) and 
0.61--0.69 keV (O\,\emissiontype{VIII}), solar wind proton flux, 
solar wind alpha flux, O$^{7+}$/O$^{6+}$, and O$^{8+}$/O$^{6+}$ 
as function of DOY in 2013. 
The vertical errors are 1$\sigma$ significance. 
The sole wind proton and alpha fluxes were taken from 
WIND SWE data (black) and ACE SWEPAM data (red). 
The oxygen ion ratios were taken from ACE SWICS data. 
The solar wind data are not corrected for the traveling time 
between the L$_1$ point to the near-Earth.
}
\label{fig9} 
\end{figure}

In figure \ref{fig9}, we plot the XIS1 0.52--0.6 keV (O\,\emissiontype{VII}) 
and 0.61--0.69 keV (O\,\emissiontype{VIII}) light curves and the solar wind proton and 
alpha fluxes taken from the WIND SWE and ACE SWEPAM data.\footnote{$\langle$ftp://spdf.gsfc.nasa.gov/pub/data/wind/swe/$\rangle$.}$^,$\footnote{$\langle$http://www.srl.caltech.edu/ACE/ASC/level2/lvl2DATA\_SWEPAM.html$\rangle$.}
These solar wind monitoring satellites orbit around the Lagrangian point L$_1$ between 
the Sun and Earth. 
The average positions of the WIND and ACE satellites during this observation were 
(259, 32) and (223, $-$12) Earth radii (R$_\mathrm{E}$) in the GSE XY plane, respectively. 
The solar wind data were not corrected for the traveling time between the L$_1$ point 
to the near-Earth. 
The solar wind proton and alpha fluxes suddenly increase just before DOY 104, which 
corresponds to the arrival of the CME-induced interplanetary shock and the passage of 
the turbulent sheath. 
The solar wind proton flux gradually decreases at the arrival of the CME, while the solar 
wind alpha flux increases again around DOY 104.8. 
The presence of enhanced alpha particles to proton ratios is rare in the ambient solar 
wind and typically associated with CMEs \citep{ric04}. 
Therefore, the solar wind alpha flux become a good indicator of the geocoronal 
SWCX events, especially associated with the CMEs.

\begin{figure}[t] 
\begin{center} 
\includegraphics[width=1.0\textwidth]{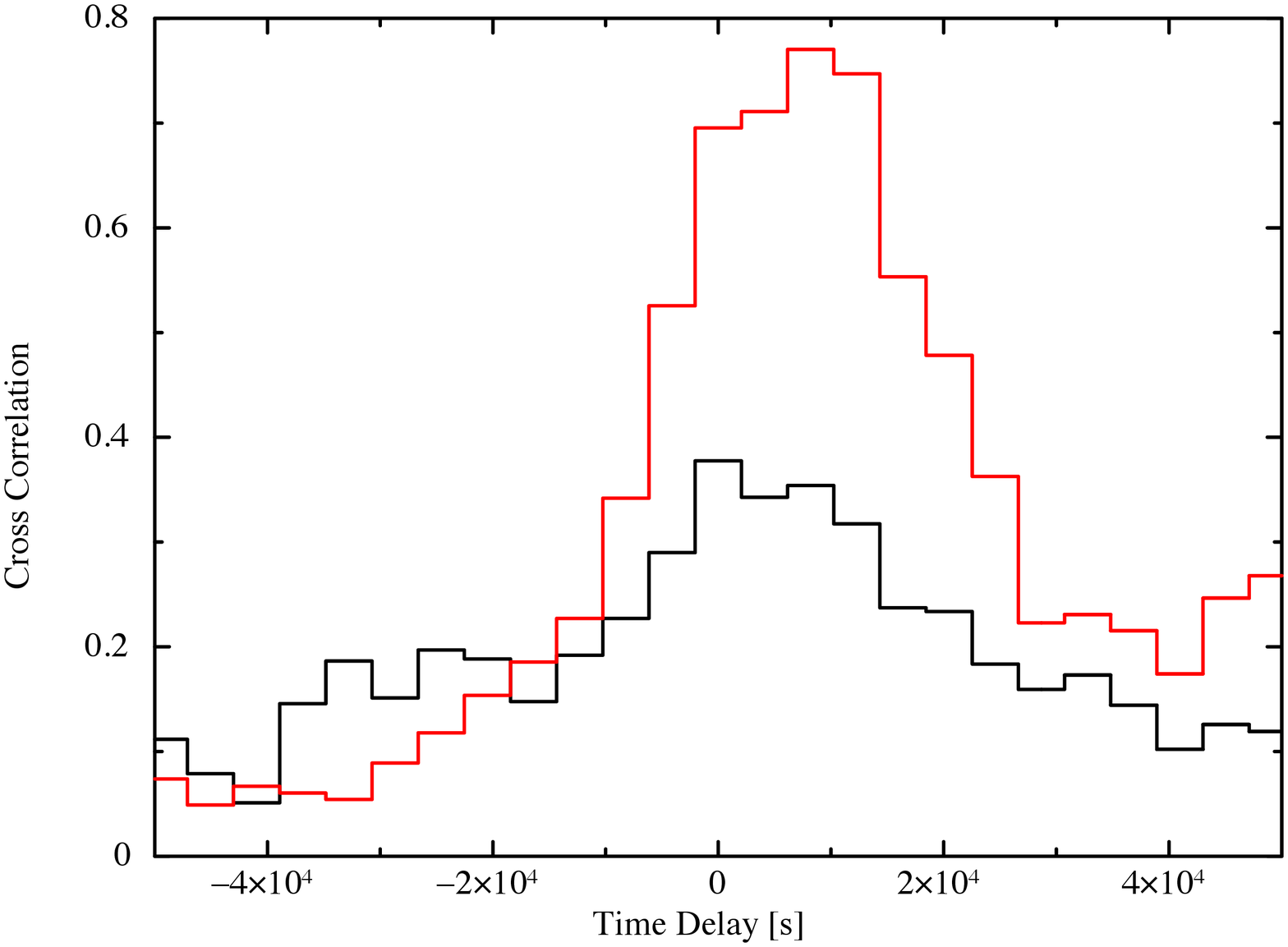} 
\end{center}
\caption{
Cross correlations between the XIS1 light curves and the WIND alpha flux. 
Black and red bars indicate the 0.52--0.6 keV (O\,\emissiontype{VII}) and 
0.61--0.69 keV (O\,\emissiontype{VIII}) correlation coefficients, respectively.
A positive delay means that the WIND data leads the XIS1 count rates.
}
\label{fig10} 
\end{figure}

To check the correlation between the geocoronal SWCX emission and 
the solar wind alpha flux, we conducted a cross-correlation analysis by using a method 
described in \citet{ezo10}. 
The ACE data was unavailable during a part of this observation. 
For this procedure, we binned the O\,\emissiontype{VII} and O\,\emissiontype{VIII} 
count rates and the WIND alpha flux into the same time bins of 4096 s. 
Figure \ref{fig10} shows the cross correlations between the X-ray light curves and 
the WIND alpha flux. 
The correlation coefficients for O\,\emissiontype{VII} and O\,\emissiontype{VIII} have 
a weak peak of $\sim$0.3 and a strong peak of $\sim$0.7, respectively. 
These peaks have a time delay of 0--12288 s. 
The positive value means that the WIND alpha flux has a time delay against the X-ray 
light curves. 
The expected time delay was estimated as $\sim$3600 s from the distance between 
the WIND and Suzaku satellites and the average solar wind speed during this 
observation. 
We then conducted the same procedure with low-resolution OMNI data time-shifted 
to the near-Earth.\footnote{$\langle$https://omniweb.gsfc.nasa.gov/ow.html$\rangle$.}
However, the OMNI alpha flux remained a time delay at 0--8192 s. 
We also noticed that the time delay for O\,\emissiontype{VIII} is slightly shifted to the 
positive side than that for O\,\emissiontype{VII}.
These time delay may be due to the solar wind transportation in the magnetosphere 
and/or the ion distribution within the CME.

\begin{figure}[t] 
\begin{center} 
\includegraphics[width=1.0\textwidth]{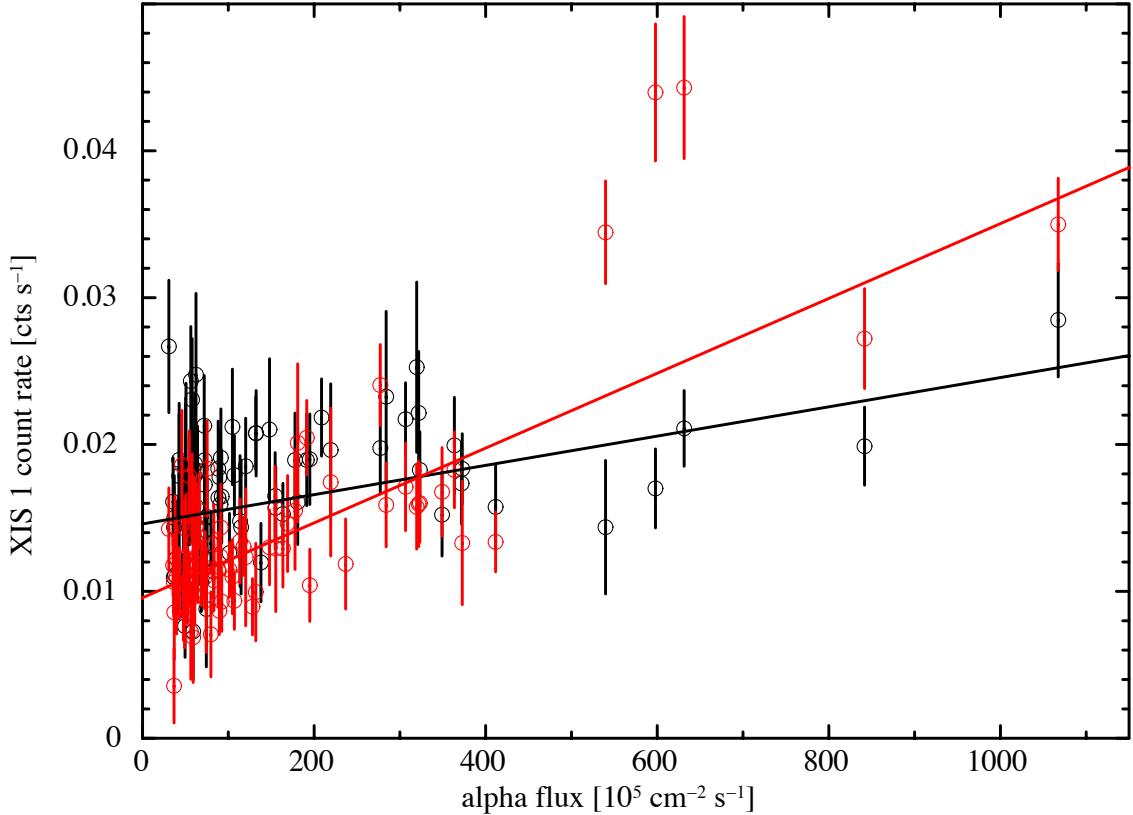} 
\end{center}
\caption{
Correlations between the XIS1 light curves and the WIND alpha flux, 
considering the no time delay (O\,\emissiontype{VII}) and the 8192 s 
time delay (O\,\emissiontype{VIII}). 
Black and red data points indicate the 0.52--0.6 keV (O\,\emissiontype{VII}) 
and 0.61--0.69 keV (O\,\emissiontype{VIII}) count rates, respectively.
The vertical errors are 1$\sigma$ significance.
The solid curves are the best-fit linear functions.
}
\label{fig11} 
\end{figure}

We simply assumed no time delay and a time delay of 8192 s to 
the WIND alpha flux corresponding to the O\,\emissiontype{VII} and 
O\,\emissiontype{VIII} count rates, respectively. 
Figure \ref{fig11} shows the correlations between the O\,\emissiontype{VII} and 
O\,\emissiontype{VIII} count rates and the WIND alpha flux considering the above 
time delay. 
These correlations are represented by a linear function expressed as 
\begin{eqnarray}
C_\mathrm{XIS} \ [\mathrm{cts} \ \mathrm{s}^{-1}] =  a \times C_\mathrm{alpha} \ 
[10^5 \ \mathrm{cm}^{-2} \ \mathrm{s}^{-1}] + b, 
\label{eq1} 
\end{eqnarray}
where $a$ is the SWCX emissivity and $b$ is an offset emission due to the instrument 
and sky backgrounds.
The O\,\emissiontype{VII} best-fit parameters are 
$a$ $=$ (1.0$\pm$0.2)$\times$10$^{-5}$ and 
$b$ $=$ (14.6$\pm$0.4)$\times$10$^{-3}$ with $\chi^2$$/$d.o.f $=$ 121.24$/$79, 
while the O\,\emissiontype{VIII} best-fit ones are 
$a$ $=$ (2.5$\pm$0.2)$\times$10$^{-5}$ and 
$b$ $=$ (9.6$\pm$0.4)$\times$10$^{-3}$ with $\chi^2$$/$d.o.f = 141.01$/$79. 
From the area of the TDX region and the spectral fitting shown in figure \ref{fig7}, 
the X-ray count rate is converted into the X-ray flux per solid angle. 
The emissivity for O\,\emissiontype{VII} and O\,\emissiontype{VIII} are 
0.009$\pm$0.002 and 0.030$\pm$0.002 LU per 10$^5$ cm$^{-2}$ s$^{-1}$, 
where LU is photons s$^{-1}$ cm$^{-2}$ str$^{-1}$. 
The positive intercepts in the O\,\emissiontype{VII} and O\,\emissiontype{VIII} 
linear functions are 13.3$\pm$0.4 and 11.2$\pm$0.4 LU, respectively. 
This means that there remains a count rate because of the XIS instrumental 
background and the sky background even if the solar wind alpha flux is zero. 
The same intercept was observed in our past geocoronal SWCX studies, 
in which the solar wind proton or ion fluxes were used to derive the coefficients 
as in \citet{ezo11} and \citet{ish13}.

In figure \ref{fig9}, we also plot the oxygen ion ratios taken from 
the ACE SWICS data.\footnote{$\langle$http://www.srl.caltech.edu/ACE/ASC/level2/lvl2DATA\_SWICS\_2.0.html$\rangle$.}
The ion densities of the solar wind are not available after 2011 August 23. 
The increased alpha particles within the CME are expected to be accompanied by
an increase in all the minor ions. 
The O$^{7+}$$/$O$^{6+}$ ratio strongly increases at DOY 104.8, while the 
O$^{8+}$$/$O$^{6+}$ ratio clearly late increases at DOY105. 
This supports that there is the time delay between the best O\,\emissiontype{VII} 
and O\,\emissiontype{VIII} coefficient peaks shown in figure \ref{fig10}. 
In addition, the peak of the enhanced O$^{8+}$$/$O$^{6+}$ ratio indicates 
that the beginning of the strong emission lines is probably delayed from 
the peak of the increased solar wind alpha flux or the arrival of the CME. 
However, the O\,\emissiontype{VIII} count rate begins to increase at DOY 104.8. 
We then checked the count rates originating from other SWCX emission lines 
(Ne\,\emissiontype{IX}, Ne\,\emissiontype{X}, Mg\,\emissiontype{XI}, 
Mg\,\emissiontype{XII}, and Si\,\emissiontype{XIII}) and found that these
count rates also increase at DOY 104.8. 
We also noticed that the O\,\emissiontype{VIII} emission line is prominent even though 
the O$^{8+}$/O$^{6+}$ ratio is smaller than the O$^{7+}$/O$^{6+}$ ratio during the 
``flare'' period. 
The sparse solar wind data hinders us to investigation a more accurate comparison. 
Therefore, the enhanced oxygen ion ratio are probably not sufficient to explain the 
beginning of the soft X-ray enhancement and the geocoronal SWCX flux against the 
ion abundances during this observation.
 
\section{Conclusion}

In this paper, we have investigated the geocoronal SWCX event 
associated with the CME on 2013 April 14--15 with the Suzaku 
data, following the same procedure as \citet{ezo11}.
We divided the data into the ``stable'', ``pre-flare'', and ``flare'' periods 
based on the characteristic features of two X-ray light curves. 
The 0.5--0.7 keV count rate increased slightly during the ``pre-flare'' 
period and by a factor of $\sim$2 during the ``flare'' period. 
The 2.5--5.0 keV count rate were almost constant during this observation.
The spectral change during the ``pre-flare'' period probably originated from the oxygen 
emission lines and that during the ``flare'' period consisted of the series of the emission 
lines from highly ionized carbon to silicon.
The former spectrum could be fitted with the two Gaussian model in 0.5--0.7 keV, 
while the later one was well represented by the Bodewits's SWCX model, the single 
Gaussian model around 0.25 keV, and the eleven Gaussian model above 0.7 keV.
The SWCX line flux ratios to O\,\emissiontype{VIII} showed a similar tendency to 
those from XMM-Newton data taken in late 2001 October 
\citep{car10}. 
From the solar wind signatures, we concluded that the ``pre-flare'' period was related 
to the CME-induced interplanetary shock and the turbulent sheath and the ``flare'' 
period was affected by the arrival of the CME. 
The highly compressed solar wind ions in the turbulent sheath resulted in the soft 
X-ray enhancement and the highly charged ions within the CME constructed the 
enriched spectrum.
However, the solar wind proton flux showed no significant time correlation. 
We thus studied the time correlation with the solar wind alpha flux and found a good 
correlation. 
The correlations between the O\,\emissiontype{VII} and O\,\emissiontype{VIII} count 
rates and the solar wind alpha flux showed a positive correlation well represented by 
the linear functions. 
This suggests that the solar wind proton flux is not always a good indicator of the 
geocoronal SWCX events associated with the CMEs and the solar wind alpha flux 
should be also investigated when the CME-induced soft X-ray enhancement is 
detected in astronomical observations.


\begin{thebibliography}{}

\bibitem[Bodewits et al.(2007)]{bod07}
Bodewits, D., et al. 2007,  A\&A, 469, 1183

\bibitem[Burlaga et al.(1981)]{bur81}
Burlaga, L., Sittler, E., Mariani, F.,  \& Schwenn, R. 1981, J. Geophys. Res., 86, 6673

\bibitem[Carter \& Sembay(2008)]{car08}
Carter, J. A., \& Sembay, S. 2008, A\&A, 489, 837

\bibitem[Carter et al.(2010)]{car10}
Carter, J. A., Sembay, S., \& Read, A. M. 2010, MNRAS, 402, 867

\bibitem[Carter et al.(2011)]{car11}
Carter, J. A., Sembay, S., \& Read, A. M. 2011, A\&A, 527, A115

\bibitem[Cravens (2000)]{cra00}
Cravens, T. E. 2000, ApJ, 532, L153

\bibitem[Cravens et al.(2001)]{cra01}
Cravens, T. E., Robertson, I. P., \& Snowden, S. L. 2001, J. Geophys. Res., 106, 24883

\bibitem[Ezoe et al.(2010)]{ezo10}
Ezoe, Y., Ebisawa, K., Yamasaki, N. Y., Mitsuda, K., Yoshitake, H.,Terada, N., Miyoshi, Y., \& Fujimoto, R. 2010, PASJ, 62, 981

\bibitem[Ezoe et al.(2011)]{ezo11}
Ezoe, Y., Miyoshi, Y., Yoshitake, H., Mitsuda, K., Terada, N., Oishi, S., \& Ohashi, T. 2011, PASJ, 63, S691

\bibitem[Fujimoto et al.(2007)]{fuj07}
Fujimoto, R., et al. 2007, PASJ, 59, S133

\bibitem[Gonzalez et al.(1994)]{gon94}
Gonzalez, W. D., Joselyn, J. A., Kamide, Y., Kroehl, H. W., Rostoker, G., Tsurutani, B. T., \& Vasyliunas, V. M. 1994, J. Geophys. Res., 99, 5771

\bibitem[Ishikawa et al.(2013)]{ish13}
Ishikawa, K., Ezoe, Y., Miyoshi, Y., Terada, N., Mitsuda, K., \& Ohashi, T. 2013, PASJ, 65, 63

\bibitem[Koyama et al.(2007)]{koy07}
Koyama, K., et al. 2007, PASJ, 59, S23

\bibitem[Koutroumpa et al.(2007)]{kou07}
Koutroumpa, D., Acero, F., Lallement, R., Ballet, J., \& Kharchenko, V. 2007, A\&A, 475, 901

\bibitem[Mckenzie et al.(1982)]{mck82}
Mckenzie, D. L., Rugge, H. R., \& Charles, P. A. 1982, J. Atomos. Terr. Phys., 44, 499 

\bibitem[Mitsuda et al.(2007)]{mit07}
Mitsuda, K., et al. 2007, PASJ, 59, S1

\bibitem[Miyoshi \& Kataoka(2005)]{miy05}
Miyoshi, Y., \& Kataoka, R. 2005, Geophys. Res. Lett., 32, L21105

\bibitem[Priest \& Forbes(2002)]{for02}
Priest, E. R., \& Forbes, T. G. 2002, A\&A Rev., 10, 313

\bibitem[Richardson \& Cane(2004)]{ric04}
Richardson, I. G., \& Cane, H. V. 2004, J. Geophys. Res., 109, A09104

\bibitem[Robertson et al.(2006)]{rob06}
Robertson, I. P., Collier, M. R., Cravens, T. E., \& Fok, M.-C., 2006, J. Geophys. Res., 111, A12105

\bibitem[Sekiya et al.(2014)]{sek14}
Sekiya, N., Yamasaki, N. Y., Mitsuda, K., \& Takei, Y., PASJ Lett., 66, L3

\bibitem[Snowden et al.(1994)]{sno94}
Snowden, S. L., McCammon, D., Burrows, D. N., \& Mendenhall, J. A. 1994, ApJ, 424, 714

\bibitem[Snowden et al.(2004)]{sno04}
Snowden, S. L., Collier, M. R., \& Kuntz, K. D. 2004, ApJ, 610, 1182

\bibitem[Wargelin et al.(2004)]{warg04}
Wargelin, B. J., Markevitch, M., Juda, M., Kharchenko, V., Edger, R., \& Dalgarno, A. 2004, ApJ, 607, 596

\bibitem[Warren \& Hughes(2004)]{warr04}
Warren, J. S., \& Hughes, J. P. 2004, ApJ, 608, 261

\bibitem[Webb \& Howard(2012)]{web12}
Webb, D. F., \& Howard, T. A. 2012, Living Rev. Sol. Phys., 9, 3

\bibitem[Yamaguchi et al.(2014)]{yam14}
Yamaguchi, H., et al. 2014, ApJ, 785, L27

\bibitem[Zurbuchen \& Richardson(2006)]{zur06} 
Zurbuchen, T. H., \& Richardson, I. G. 2006, Space Sci. Rev., 123, 31

\end{thebibliography}
\end{document}